\documentclass[letterpaper,11pt]{article}
\usepackage{amsmath,amssymb,mathtools}
\usepackage{algorithm,algpseudocode}
\usepackage{booktabs}
\usepackage{physics}
\usepackage{fullpage}
\usepackage{hyperref}
\usepackage{makecell}
\usepackage{natbib}
\usepackage{palatino}
\usepackage[disable]{todonotes}  % For production 

\begin{document}

\title{Quantum and Classical Combinatorial Optimizations Applied to Lattice-Based Factorization}

\author{Willie Aboumrad, Dominic Widdows, Ananth Kaushik \\
\normalsize\texttt{aboumrad|widdows|kaushik@ionq.com} \\
IonQ Inc.
}

% Slight abuse of \date field.
\date{\today}

\maketitle

\begin{abstract}
    The availability of working quantum computers has led to several proposals and claims of quantum advantage. 
    In 2023, this has included claims that quantum computers can successfully 
    factor large integers, by optimizing the search for nearby smooth integers (numbers all of whose prime factors are small).

    This paper demonstrates that the hope of factoring numbers of commercial significance
    using these methods is unfounded. Mathematically, this is because the density of 
    smooth numbers decays exponentially as 
    $n$ grows and numerical evidence suggests lattice-based methods are not particularly well-suited for finding these.
    Our experimental reproductions and analysis show
    that lattice-based factoring does not scale successfully to larger numbers, 
    that the proposed quantum enhancements do not alter this conclusion, and that other
    simpler classical optimization heuristics perform much better for lattice-based factoring.
    
    However, many topics in this area have interesting applications 
    and mathematical challenges, independently of factoring itself. 
    We consider particular cases of the CVP, and opportunities for applying quantum techniques
    to other parts of the factorization pipeline, including the solution of linear
    equations modulo 2. Though the goal of factoring 1000-bit numbers is still out-of-reach, 
    the combinatoric landscape is promising, and warrants further research with more circumspect
    objectives.
    
\end{abstract}

%%%
\section{Introduction and Outline}

The story of lattice-based factoring goes back to the 1990's, when Schnorr published work outlining the possibilities \citep{schnorr1990factoring}. This work was published in a peer-reviewed journal, and it led to
interesting further work, including demonstrating that the Shortest Vector
Problem (SVP) is NP-hard.

However, Schnorr's method did not lead to prime factorization in practice. In 2021, \citet{schnorr2021fast} claimed to have studied a variant of the method in enough detail to claim that it would be able to factorize
integers up to 2048 bits, thus cracking the current RSA encryption protocol. 
This work has not been published in peer-reviewed venues,
but the claim remains in the preprint version.
This was built on by \citet{yan2022factoring} and \citet{hegade2023digitized} to propose quantum versions.

This article shows that neither the Schnorr
factorization claims, or the proposed quantum improvements, perform
effectively enough to factorize numbers above about $80$ bits (current hardware), and there is no reason to believe that the methods will 
provide the most effective methods for factorizing larger numbers. Indeed, we provide a purely classical heuristic that outperforms the claimed quantum advantage, but is still not nearly efficient enough
to compete with more standard sieving methods. 
Thus we conclude that factoring remains an exponentially hard problem using Schnorr's method, and given the status of current hardware, that systems based on the assumption that factoring is a difficult problem (such as RSA encryption) are not currently at risk.

The article is laid out as follows. Section \ref{fact_basics} gives a background on integer
factoring, enough to explain Schnorr's lattice-based method in Section \ref{sec:schnorrgate}.
Section \ref{sec:proposed_quantum} explains the proposed quantum optimizations, and 
Section \ref{sec:experiments} describes our experiments and results.
The paper finishes with a brief summary of other attempts to optimize the factorization process,
and other opportunities for quantum advantage in this area, including combinatoric challenges
such as the knapsack problem, and the solution of linear equations modulo 2.

%%%
\section{Factorization Basics --- Congruence of Squares}
\label{fact_basics}

Most modern factorization methods are based on Dixon's method \citep{dixon1981asymptotically}, which itself goes back to Fermat's method. The following observations underlie these methods. Say we seek to factorize an integer $n$. Then if we obtain $x, y$, with $x \not \equiv \pm y \mod n$, satisfying
\begin{equation}\label{cong_of_squares}
  x^2 \equiv y^2 \mod n, 
\end{equation}
both $\gcd(x-y, n)$ and $\gcd(x + y, n)$ are non-trivial factors of $n$, 
which may be computed efficiently using Euclid's algorithm.\footnote{See \citet[Bk VII, Prop 2]{joyce1996euclid}} 
For instance, suppose $n = 1649$, and note that $x = 1743$ and $y = 80$ satisfy 
\begin{equation*}
	x^2 \equiv 1453 \equiv y^2 \mod n.
\end{equation*}
Clearly $x \not \equiv \pm y \mod n$, so 
\[\gcd(x - y, n) = \gcd(1683, 1649) = 17,\] 
and
\[\gcd(x + y, n) = \gcd(1843, 1649) = 97\] 
are non-trivial factors of $n$. 

Thus factorization reduces to constructing a solution to congruence \eqref{cong_of_squares} satisfying $x \not \equiv \pm y \mod n$; in other words, it reduces to finding an \textit{interesting solution} to \eqref{cong_of_squares} \cite{pomerance2008smooth}. There are many \textit{uninteresting} solutions, i.e., those that also satisfy $x \equiv \pm y \mod n$, like $(x, x)$ and $(x, -x)$ for any $x \in [0, n)$, that do not lead to factorizations of $n$.
Factorization methods cannot discern interesting solutions to \eqref{cong_of_squares} a priori, but in practice this is not a significant impediment because the probability that an arbitrary solution to \eqref{cong_of_squares} is interesting is at least $50\%$: for details, review the simple enumeration argument in \cite{pomerance2008smooth}. 
%See bottom of page 70 in Smooth numbers by pomerance

Even Shor's quantum factorization algorithm \cite{shor1994algorithms} finds factors by constructing interesting solutions to \eqref{cong_of_squares}. Shor's method proceeds as follows. To begin, it randomly selects an integer $a \in (1, n)$ coprime to $n$ and then uses a quantum algorithm, based on the quantum Fourier transform, to produce an optimized order-finding subroutine \cite[\S A4.3 and \S 5.3]{nielsen2002quantumcomputation} to compute the smallest integer $r$ such that 
\begin{equation}\label{shor_cong}
	a^r \equiv 1 \mod n.
\end{equation}
If the \textit{order} $r$ of $a$ is odd, the algorithm draws a different starting value and tries again until it finds an integer with even order.\footnote{The probability that $r$ is even is greater than $1 - \frac{1}{2^m}$, where $p_m$, the $m^{th}$ prime number, is the largest prime factor of $m$ \cite[Theorem 5.3]{nielsen2002quantumcomputation}.}
Having produced a solution to congruence \eqref{shor_cong} with $r$ even, Shor's algorithm obtains a solution to \eqref{cong_of_squares} by setting $x = a^{r/2}$ and $y = 1$. 
This solution is interesting only if $a^{r/2} \not \equiv -1 \mod n$, or equivalently if $r \equiv 2 \mod 4$, and Shor's algorithm continues randomly selecting $a \in (1, n)$ until it finds an interesting solution to \eqref{cong_of_squares} via $x = a^{r/2}$ and $y = 1$.

Although factorization methods differ in how they obtain solutions to \eqref{cong_of_squares}, the construction usually involves two stages: data collection and processing. 

\subsection{Data Collection and Smooth Numbers}
\label{sec:mod2_processing}

In the data collection phase factoring methods search for certain \textit{smooth integers}. A number is said to be $p$-smooth if it has no prime factors larger than $p$. If the first $m$ prime numbers are written as $\{p_1, p_2, \ldots, p_m\}$,
then any $p_m$-smooth number can be written as a product 
$\Pi_{i=1}^{m} p_i^{e_i}$. 
This pre-determined set of ``small'' primes is often called the \textit{factor basis}. The data collection phase seeks $m+1$ $p_m$-smooth numbers. We note that the size $m$ of the factor basis is typically a hyperparameter that must be tuned.

The data collection typically involves searching a large space. For instance, the Quadratic Sieve collects smooth integers by sieving a large interval centered around $\sqrt{n}$;
the General Number Field Sieve, currently the fastest known method for factorizing integers with more than $100$ digits, refines the search by exploiting the theory of algebraic number fields \citep{pomerance1996tale}. 
The collection step accounts for the bulk of the computational load: in the record-breaking RSA-$250$ factorization completed in February $2020$, the sieving step accounted for roughly $2,450$ of the $2,700$ core-years (using a single Intel Xeon Gold $6130$ CPU running a $2.1$ GHz clock rate as a reference) \citep{boudot2022state}. 

\subsection{Processing Relations from Smooth Numbers}

In the processing step, the methods combine several congruences to produce a solution to \eqref{cong_of_squares}. 
The technique goes back to the work of Maurice Kraitchik in the 1920s, as described by \cite{pomerance1996tale}.
For example, to factor 2041, the smallest number $x$ such that $x^2 > 2041$ is 46, and we have
\[ 46^2 \equiv 75, \quad 47^2 \equiv 168, \quad 49^2 \equiv 360, \quad 51^2 \equiv 560 \quad \mathrm{(all\ mod\ 2041).} 
\]
These remainders all have prime factors no greater than 7:
\[ 75 = 3 \cdot 5^2, \quad 168 = 2^3 \cdot 3 \cdot 7, 
\quad 360 = 2^3 \cdot 3^2 \cdot 5, \quad 560 = 2^4 \cdot 5 \cdot 7.
\]
It is easy to multiply these by adding the exponents, and to deduce that
\[ (46\cdot 47\cdot 49 \cdot 51)^2 \equiv 2^{10} \cdot 3^4 \cdot 5^4 \cdot 7^2 \pmod{2041}.
\]
From here, we deduce that $x = 46\cdot 47\cdot 49 \cdot 51 \equiv 311 \pmod{2041}$ 
and $y = 2^5 \cdot 3^2 \cdot 5^2 \cdot 7 \equiv 1416 \pmod{2041}$ give a solution to \eqref{cong_of_squares}.
Lastly we compute $\gcd(1416-311, 2041) = 13$, and indeed $2041 = 13 \times 157$.

Finding relations where the remainder modulo $n$ is smooth is crucial,
because it allows for the use of linear algebraic techniques via \textit{index calculus}, as the following example illustrates. Suppose we fix $P = \{p_1, \ldots, p_m\}$ as the factor basis, with $p_j$ denoting the $j^{\mathrm{th}}$ prime.
In addition, suppose that in the data collection step our method found $m + 1$ integers $x_i$, for $i = 1, \ldots, m + 1$, such that $x_i^2 \mod n$ is $p_m$-smooth. This means there exist $\Vec{e}_i \in \mathbb{Z}_{\geq 0}^{m}$ such that 
\begin{equation}\label{square_as_prime_prod}
	x_i^2 \equiv p_1^{e_{i1}} \cdots p_m^{e_{im}} \mod n.
\end{equation}

Using multi-index notation, we may write $x_i^2 \equiv p^{\Vec{e}_i} \mod n$. Note that \eqref{square_as_prime_prod} implies 
\begin{equation*}
	x_i^2 \cdot x_k^2 \equiv \prod_{j=1}^m p_j^{e_{ij} + e_{kj}} \mod n,
\end{equation*}
or equivalently $x_i^2 \cdot x_k^2 \equiv p^{\Vec{e}_i + \Vec{e}_k} \mod n$. Thus for any subset $\mathcal{I} \subseteq \{1, \ldots, m+1\}$, the product $\prod_{i \in \mathcal{I}} x_i^2 = \big(\prod_{i \in \mathcal{I}} x_i\big)^2$ is a square that is equivalent to $p^{\sum_{i \in \mathcal{I}} \Vec{e}_i}$ modulo $n$. Therefore if $\mathcal{I}$ is such that $p^{\sum_{i \in \mathcal{I}} \Vec{e}_i}$ is also a square, we obtain a solution to \eqref{cong_of_squares} via 
\[
	x = \prod_{i \in \mathcal{I}} x_i
	\quad\text{and}\quad
	y = p^{\frac{1}{2}\sum_{i \in \mathcal{I}} \Vec{e}_i}.
\]

We may construct such a subset $\mathcal{I}$ by solving a linear system. Since
$$
	p^{\sum_{i \in \mathcal{I}} \Vec{e}_i} = p^{\sum_{i = 1}^{m+1} z_i \Vec{e}_i}
$$
with $z_i = 1$ if $i \in \mathcal{I}$ and $z_i = 0$ otherwise, $p^{\sum_{i = 1}^{m+1} z_i \Vec{e}_i}$ is a square if and only if each coordinate of $\sum_{i = 1}^{m+1} z_i \Vec{e}_i$ is even. Thus we may construct $\mathcal{I}$ by finding $z_1, \ldots, z_{m+1} \in \mathbb{F}_2$ such that 
\begin{equation}
	\sum_{i = 1}^{m+1} z_i \Vec{e}_i \equiv 0 \mod 2.
\end{equation}
In other words, we may obtain the desired index set $\mathcal{I}$ by reducing $E = 
\begin{bmatrix}
	\Vec{e}_1 & \cdots & \Vec{e}_{m+1} 
\end{bmatrix}
$ 
modulo $2$ and solving the homogeneous system $E z = 0$. 
Note that a non-trivial solution is guaranteed to exist when $E$ has more columns than rows, i.e., when we have collected at least one more smooth number than there are primes in the factor basis.

There is a clear algorithmic trade-off here: a smaller smaller factor basis $\{p_i : i = 1\ldots m\}$ reduces both the number of relations needed and the complexity of deducing a solution to \eqref{cong_of_squares} from these relations, but it makes it harder to
find candidate relations where the remainder mod $n$ is $p_m$-smooth.

% The use of negative factors is an extra detail here: starting the prime basis with $-1$ (using $\{-1, 2, 3, \ldots \}$ instead of $\{2, 3, 5, \ldots\}$ as a factor basis) enables us to include auxiliary that are slightly smaller (as well as slightly larger) than $n$. For example, $45^2 \equiv -16 \pmod{2041}$, which can be written as 
% $45^2 \equiv p_0^1\cdot p_1^4 \pmod{2041}$ with the factor basis $\{p_0=-1, p_1=2\}$.
% The rest of the solution process remains exactly the same, so we can effectively double the number of 
% smooth relations we can find, at the cost of adding one new element to the factor basis.

%%%
\section{Schnorr's Lattice-Based Factoring and the Closest Vector Problem}
\label{sec:schnorrgate}

We are now in a position to explain  in detail the data collection and processing subroutines defined by Schnorr's method.
In \cite{schnorr1990factoring}, Schnorr claimed to have devised a method that can produce smooth integers useful for factorization, by computing approximate solutions to instances of the Closest Vector Problem (CVP).

\subsection{From Factoring to the Closest Vector Problem}

The connection between integer factoring and the CVP
problems lies in the data collection phase. We use logarithms to 
translate the problem of finding small integers whose product is close to $n$, into the problem of 
finding combinations of small integers with logarithms that sum to the logarithm of $n$.
The linearity of this formulation means that the set of possible combinations 
can be described as a lattice of vectors (whose dimension is given by the size of the factoring basis).
(In this context, a {\it lattice} is the set of all integer linear combinations of a given set of basis vectors,
so for a basis $B = \{b_1, \dots, b_m\}$, the \textit{lattice generated by} $B$ is $\{\lambda_1 b_1 + \ldots + \lambda_m b_m : \lambda_i \in \mathbb{Z}\}$.)
The search for useful approximate factoring relations is then formulated as the search for 
vectors in this lattice that are close to a target vector which is determined by $\ln n$.
In essence, this reduces the data collection step of the prototypical factorization algorithm to a sequence of combinatorial optimization problems. 

Schnorr's method is based on the following observations. Again, suppose we seek to factor an integer $n$, set $p_0 = -1$, and let $p_1, \ldots, p_m$  denote the first $m$ primes. In addition, suppose we obtain integers $e_j$ such that
\begin{equation}\label{log_approx}
	\epsilon \coloneqq \bigg|\sum_{j=1}^m e_j \ln p_m - \ln n \bigg| \approx 0.
\end{equation}
Thus if we set 
\begin{equation}\label{smooth_pair_defn}
	u = \prod_{e_j \geq 0} p_j^{e_j}
	\quad \text{and} \quad
	v = \prod_{e_j < 0} p_j^{-e_j}, 
\end{equation}
we obtain 
\begin{equation*}
	\bigg|\ln \bigg(\frac{u}{vn}\bigg)\bigg| = \epsilon, \\
	\quad\text{which means}\quad
	u - vn = vn(e^\epsilon - 1) \approx \epsilon vn.
\end{equation*}
The last approximation follows from Taylor's theorem. Shnorr's claim is that since $\epsilon$ is small, $u - vn$ is also small and therefore it is \textit{likely} to be $p_m$-smooth. Such pairs of integers play a key role: 
Schnorr calls a pair $(u, v)$ such that both $u$ and $u - vn$ are $p_m$-smooth a \textit{fac-relation}; in \cite{yan2022factoring} fac-relations are known as \textit{smooth relation pairs}.

Schnorr's method produces the coefficients in \eqref{log_approx} by approximating solutions to the CVP, in order to obtain smooth pairs $(u, v)$ as in \eqref{smooth_pair_defn}. 
Many different CVP instances are considered, by randomly assigning different weights to different elements 
of the factor basis.
In particular: let $B_{m, c} \in \mathbb{Z}^{(m+1) \times m}$ denote the \textit{prime lattice} generated by the columns of
\begin{equation}
\label{prime_lattice}
B_{m, c} = 
\begin{bmatrix}
	f(1) & 0 & 0 & 0 \\
	0 & f(2) & 0 & 0 \\
	\vdots & \vdots & \ddots & \vdots \\
	0 & 0 & \cdots & f(m) \\
	\lceil 10^c \ln(p_1) \rceil & \lceil 10^c \ln(p_2) \rceil & \cdots & \lceil 10^c \ln(p_m) \rceil
\end{bmatrix},
\end{equation}
where $c$ is a tunable parameter,\footnote{Papers are often vague on what constant $c$ should be used:
\citep[\S 4]{schnorr1990factoring} uses values between 2 and 3, and \cite{yan2022factoring} use numbers ranging at least
from 1 to 4, but the choice is not explained.} 
and $f(j) = \log p_{\sigma(j)}$, with $\sigma$ denoting a random permutation on $m$ letters. 
%We note that Relation \ref{default_params} describes the choice of tunable parameters used by default by our implementation of Schnorr's method. % removed this, it's clear later
Now let $t \in \mathbb{Z}^{m+1}$ denote the \textit{target vector}
\begin{equation*}
	t = 
\begin{bmatrix}
	0 \\
	\vdots \\
	0 \\
	\lceil 10^c \ln n \rceil
\end{bmatrix}.
\end{equation*}
The point is that if $e \in \mathbb{Z}^m$ is such that $B_{m, c}e$ is the prime lattice vector closest to $t$, the difference in \eqref{log_approx} is as small as possible. This gives the corresponding $u - vn$ a better chance of being smooth, 
amongst most other candidates corresponding to different prime lattice vectors. It is difficult to quantify the smoothness probability of $u - vn$ as a function of its defining prime lattice vector, because it depends on both the estimation error $\epsilon$ and the smooth number $v$, and these are interrelated. 
Regardless, we note that the amongst prime lattice points that are equidistant to the target vector $t$, the smoothness probability of $u - vn$ is largest for the pair with smallest $v$; in other words, prime lattice points with smaller negative coordinates yield higher quality solutions.

While Schnorr's original work derives some performance guarantees, in the form of $\ell_\infty$- and $\ell_1$-norm bounds, they are asymptotic and rely on norms that are irrelevant to the CVP heuristic algorithms employed in practice.
In particular, Schnorr proves that if we find $e \in \mathbb{Z}^m$ satisfying
\begin{align*}
	||B_{m, c}e - t||_\infty &\leq \ln p_m, 
	\quad\text{and} \\
	||B_{m, c}e - t||_1 \,\, &\leq (2c - 1)\ln n + o(\ln p_m)
\end{align*}
then we have that $|S(u, v)| = p_m^{o(1)}$, with the asymptotic for $n \to \infty$ \cite[Lemma~2]{schnorr1990factoring}. The conclusion suggests that if we find $e$ such that $B_{m, c}e$ is sufficiently close to $t$, then $S$ is sufficiently likely to be smooth, because it is a sufficiently small integer. 
In summary, \citet{schnorr1990factoring} argued that if we can solve the CVP, 
we can use this to find lots of useful fac-relations, and
solve the factorization problem.

Perhaps the most contentious point is the size of the lattice needed in order to reliably collect smooth relations: numerical evidence from our experiments, described in Section \ref{sec:experiments}, suggests both \citet{schnorr1990factoring} and \citet{yan2022factoring} vastly underestimate it. 

\begin{figure}
    \centering
    \includegraphics[width=4in]{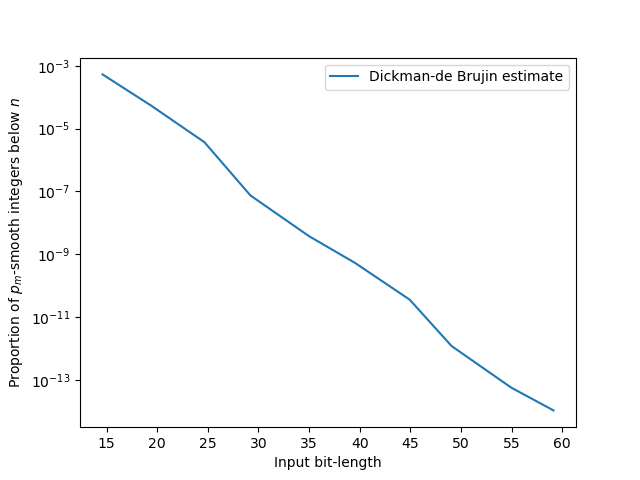}
    \caption{When the lattice dimension $m = \log(n) / \log(\log(n))$ is sublinear in the bit-length of $n$, the proportion of $p_m$-smooth integers available for collection is exponentially small.}
    \label{fig:dickman_estimate}
\end{figure}

As a guideline, we use number theoretic results developed between the 1930s and 1950s to estimate the density of smooth numbers below $n$ as $n$ gets large
\citep{ramaswami1949number,debruijn1951number}.
Concretely, we combine Dickman's function with the Prime Number Theorem to estimate that the proportion of smooth integers below $n$ available for collection under \citet{yan2022factoring}'s regime of sublinear lattice dimension is exponentially small, as follows. To begin, let $\Psi(n, p)$ denote the number of $p$-smooth integers below $n$. Then a theorem of Dickman guarantees that for fixed $a$, $\Psi(n, n^{1/a}) \sim n \rho(a)$, with $\rho$ denoting the Dickman (or Dickman-de Brujin) function. 
This is the continuous function satisfying the differential equation 
\begin{equation}\label{dickman_fun}
x \rho'(x) + \rho(x - 1) = 0    
\end{equation}
with initial conditions $\rho(x) = 1$ for $0 \leq x \leq 1$. Therefore,
$$
\frac{\Psi(n, n^{1/a})}{n} \sim \rho(a);
$$
the proportion of $n^{1/a}$-smooth integers below $n$ grows like $\rho(a)$. Now the Prime Number Theorem estimates that
$p_m \sim m \log m$, so it follows that
$$
\frac{\Psi(n, p_m)}{n} \sim \rho\bigg(\frac{\log(n)}{\log(m) + \log(\log(m))}\bigg),
$$
since $n^{1/a} = m \log(m)$ when $a = \log(n) / (\log(m) + \log(\log(m))$. The plot in Figure \ref{fig:dickman_estimate} shows that when $m = \log(n) / \log(\log(n))$ is sublinear in the bit-length of $n$, as \citet{yan2022factoring} suggest, the proportion of smooth integers below $n$ available for collection is exponentially small.

\subsection{Vector Lattices and Approximate Solutions to the Closest Vector Problem}

\begin{figure}
    \centering
    \includegraphics[width=5in]{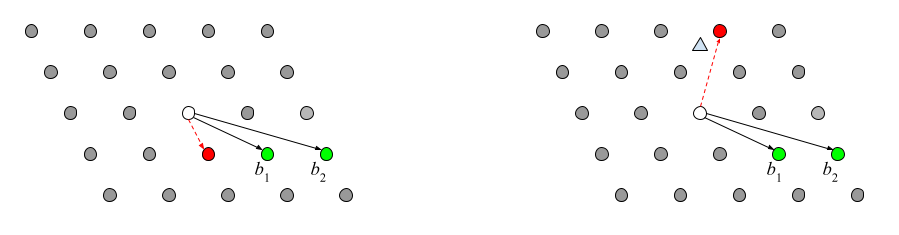}
    \caption{Shortest and Closest Vector Problems. For the SVP (left), the task is to find the shortest non-zero lattice vector 
    (integer linear combination of $b_1$ and $b_2$).
    For the CVP (right), the task is to find the lattice vector that is closest to the target blue triangle.}
    \label{fig:svp-cvp}
\end{figure}

Given a lattice, the Shortest Vector Problem (SVP) is to find the shortest nonzero vector in the lattice,
as measured by some norm. The Closest Vector Problem (CVP) is to find the closest lattice vector to a given target vector (Figure \ref{fig:svp-cvp}).
Both of these problems also have approximate alternatives, for example, to find any vector whose distance
is no more than $1 + \varepsilon$ times the minimal possible distance.
The problems are closely related, though not identical.
Generally, the CVP is considered to be the harder
problem, partly because the SVP can be solved for a basis $B = \{b_1, \ldots, b_n\}$ if a version of the CVP can be solved
for each basis vector
\citep{micciancio2001hardness,micciancio2002complexity}.

By the early 1980s, \citet{van1981another} showed that the CVP is NP-complete, and the SVP is NP-hard with the $l_\infty$ norm. By the late 1990s, many more results were known, including that that the SVP with the $l_2$ norm is NP-hard 
(for randomized reductions, that is, there is a probabilistic Turing-machine that
reduces any problem in NP to a polynomial number of instances of the SVP) \citep{ajtai1998shortest}. As \citet{ajtai1998shortest} indicates,
the potential application to factoring integers, arising from  
\citet{schnorr1990factoring} and related work, was one of the motivations
for these research efforts.

Particularly important constructions in vector lattices, discovered during the 1980s, include the Lenstra–Lenstra–Lovász (LLL) lattice basis reduction algorithm, and Babai's nearest plane algorithm. 

The LLL algorithm reduces a basis to an equivalent basis with short, nearly orthogonal vectors, in time polynomial in the basis length \citep{lenstra1982factoring}. It has become a standard simplification step, and the algorithm
is supported in several software packages.

Babai's nearest plane algorithm is explained in \cite{babai1986lovasz}. Given an LLL-reduced basis for a lattice, it finds a vector
whose distance from the target vector $t$ is not greater than a factor of $2(\frac{2}{\sqrt{3}})^d$ times 
the closest possible distance, for a lattice of dimension $d$.
Let $b_1^*, \ldots, b_n^*$ denote an LLL-reduced basis (sorted by increasing norm) for the prime lattice $B_{n, c}$. 
Additionally, let $G = \begin{bmatrix} g_1 & \cdots & g_n\end{bmatrix}$ denote the matrix obtained by running Gram-Schmidt orthogonalization (without normalizing the columns) on $b_1^*, \ldots, b_n^*$. 
Then Babai's nearest plane algorithm essentially suggests we should calculate the projection of the target vector onto the span of $G$, round the resulting coefficients to the nearest integer, and use those coefficients to return a linear combination of the LLL-reduced basis.
The algorithm computes the coefficients sequentially. First set $b ^{\mathrm{opt}} \leftarrow t$. Then for each $j$ from $m$ down to $1$, compute $\mu_j = \langle t, g_j \rangle / \langle g_j, g_j \rangle$, let $c_j = \lceil \mu_j \rfloor$ denote the nearest integer to $\mu_j$, and update 
\begin{equation}\label{babai_update}
    b^{\mathrm{opt}} \leftarrow b^{\mathrm{opt}} - c_j b_j^{.*}
\end{equation}
The name `Babai's nearest {\it plane} algorithm' is used because at step $j$, we take the plane or more general $j$-dimensional 
subspace 
$\mathrm{Span}\{g_1, \ldots, g_j\}$,
and find the integer $c_j$ such that the distance from $c_j b_j + \mathrm{Span}\{g_1, \ldots, g_{j-1}\}$ to the target vector $t$ is minimized. 
So each step finds a nearest (hyper)plane, and the algorithm's output (the final $b^{opt}$) is a vector close to the target vector $t$.

Key theoretical results and heuristic techniques for working with vector lattice problems were thus developed during the 1980s and 1990s, 
during the same period that the main factoring approaches we use today were established (c.f. \citep{dixon1981asymptotically} -- \cite{pomerance1996tale}).
When \citet{schnorr1990factoring} proposed that this application of the CVP could be the key step to enable 
polynomial-time prime factorization, only some of the complexity results and bounds that we know today were established. 
Approximating heuristics for solving the CVP account for most of the computational load in Schnorr's method.

The challenge this leaves is to demonstrate whether these heuristics are effective enough, to find close-enough 
vectors, that generate enough useful fac-relations, to create enough modular equations, so that the 
processing strategy of Section \ref{sec:mod2_processing} works to find solutions to \eqref{cong_of_squares}.
This challenge is discussed in the next section, and forms the main theme for much of the rest of this paper.

%%%
\subsection{Does the Schnorr Method Work?}

While Schnorr's method can be implemented end-to-end,
we have found no evidence that it can be used to factorize numbers
that cannot already be factorized much faster using an established sieving technique.

When introducing the method, \citet{schnorr1990factoring} claimed that the reduction of the factoring problem to the search for fac-relations worked
in polynomial time, so long as we can make heuristic assumptions about
the distribution of smooth integers. 
Much has been clarified in the theory of lattice problems since: 
for example, Lemma 2 of \citet{schnorr1990factoring} is proved 
for the $\ell_1$ and $\ell_\infty$ norms, 
which have limited practical bearing, because 
the most popular CVP heuristics rely on $\ell_2$ measurements. 
This was before \cite{ajtai1998shortest}
demonstrated the NP-hardness (under random reduction) 
of the SVP under the $\ell_2$-norm.

Schnorr has proposed extensions to these proposals,
based on optimizations including pruning \citep{schnorr2013factoring}, 
permutation, and primal-dual reduction \citep{schnorr2021fast},
though these are invited papers and preprints, not peer-reviewed publications.\footnote{An earlier preprint with
some of the material in \citet{schnorr2021fast} was retracted; we refer to the newer preprint, ``Fast Factoring Integers by SVP Algorithms, corrected''.}
Claims that the methods lead to fast factoring algorithms are based on complexity
analyses, and involve delicate considerations of many of the approximations and bounds
in results on lattices. In addition, \citep{schnorr2021fast} added publicity by adding the claim 
that ``this destroys the RSA cryptosystem'', but 
these claims have been disputed and largely dismissed, not in refereed publications,
but in online forums.\footnote{Online discussions include {\it Does Schnorr's 2021 factoring method show that the RSA cryptosystem is not secure?} at \url{https://crypto.stackexchange.com/questions/88582/}. All the answers agree that it
does not, some based on detailed algorithmic analyses, and some emphasizing that 
it's so easy to demonstrate when a factorization algorithm works for large numbers, by presenting their factors 
--- in the absence of which, we must assume that the algorithm doesn't work for large numbers.}

Code to analyze the extraction of suitable fac-relations 
at large scales has been shared by \cite{ducas2021schnorrgate},
and works
using the {\sc SageMath} platform 
\citep{sagemath}.
The motivation here is partly to investigate the complexity of the
underlying lattice operations and the accuracy of approximations, 
furthering earlier research
\citep{ducas2018shortest}.
His attempts to produce sufficient fac-relations to enable factorization
demonstrate that the algorithmic analysis of these problems is fascinating
and more difficult than we might intuitively expect, but that 
it does not perform well enough to compete with standard sieving methods.

In our own experiments
(described below), we have successfully implemented an end-to-end factorization workflow using Schnorr's methods, 
but they have been unable to factorize numbers larger than $72$ bits when running overnight, whereas the quadratic sieve
method has factorized $110$ bit and larger numbers in a matter of moments. 
The challenge is not only in finding fac-relations, but also in finding fac-relations with enough variety to solve the resulting system of equations: it happens that many different random lattice diagonal permutations result in the same fac-relation.

Nonetheless, the mathematical and computational results reviewed so far
left a slight possibility, at least in principle. 
Is there some approximate solution to the CVP, that is closer than the Babai nearest plane approximation, but does not suffer from the NP-hardness of solving the full CVP, so that the search for fac-relations is both tractable and fruitful enough to solve the factoring problem? This is the hope that
encouraged the quantum approaches to which we turn next.

%%%
\section{Proposed Quantum Optimizations}
\label{sec:proposed_quantum}

The claims of \cite{schnorr2021fast} have motivated attempts
to factorize integers using quantum computers, including
those reported in the preprints of \citet{yan2022factoring} and \citet{hegade2023digitized}. While these works have attracted some 
attention, the same caveats apply as with \citet{schnorr2021fast}: 
the methods have not demonstrated prime factors at a challenging scale,
the works are not peer-reviewed, and experts in the field have expressed
doubts and concerns \citep{aaronson2023cargo}.

The main contribution in \cite{yan2022factoring} is to use the Quantum Approximate Optimization Algorithm (QAOA, \cite{farhi2014quantum})
to refine the CVP approximations produced by Babai's method, in the 
hope that this refinement leads to more useful fac-relations.
In essence, QAOA serves as a sophisticated rounding mechanism: instead of greedily rounding each $\mu_j$ to the nearest integer in isolation in Babai's
algorithm, 
we take a holistic view and choose $x_j \in \{0, 1\}$ to minimize
\[
\norm{t - \bigg(b^{\mathrm{op}} + \sum_{j=1}^m x_j \kappa_j b_j^*\bigg)}_2^2,
\]
with $b^{\mathrm{op}}$ denoting Babai's CVP approximation, $\kappa_j = \mathrm{sign}(\mu_j - c_j)$, and $b_j^*$ denoting the $j$th column of the LLL-reduction  of $B_{m, c}$. We note that $\kappa$ is known as the \textit{coding vector} in \citet{yan2022factoring}.

Choosing a minimizing $x \in \{0, 1\}^m$ is tantamount to solving a Quadratic Unconstrained Binary Optimization (QUBO) program, which \citet{yan2022factoring} interpret as a minimum-energy eigenstate problem via the Ising map. In particular, if we let $\sigma^z_j$ denote the Pauli-$Z$ gate acting on the $j$th qubit and set $Z_j = \frac{1}{2}(I - \sigma^z_j)$, the problem Hamiltonian is given by
\begin{equation}\label{babai_hamiltonian}
    H = \sum_{i=1}^{m+1} \bigg(t_i - b^{\mathrm{op}}_i - \sum_{j=1}^m \kappa_j b_{ij}^* Z_j\bigg)^2.
\end{equation}
Since $H$ is diagonal with respect to the computational basis, we see that each eigenstate of $H$ corresponds to one of the $2^m$ possible roundings. In turn, this means that every $H$-eigenvector $\ket{\psi}$ with lower energy than $\ket{0}$ produces an enhanced CVP solution via
\begin{equation}\label{cvp_refinement}
    b^{\ket{\psi}} \coloneqq b^{\mathrm{op}} + \sum_{j=1}^m \kappa_j \psi_j b_j^*.
\end{equation}

The hypothesis in \cite{yan2022factoring} is that lower-energy eigenstates are more likely to yield smooth relation pairs, via Relation \ref{smooth_pair_defn} with the lattice coordinates $e$ defined by
\begin{equation}\label{lattice_coords}
    b^{\ket{\psi}} = B_{m, c} e.
\end{equation}
They explain in detail some of the steps to factorize an $11$-bit, a $26$-bit, and a $48$-bit
number, using $3$, $5$, and $10$ qubits respectively. The number of qubits here corresponds to the dimension of the lattice used. 

Following \citet{yan2022factoring}, a preprint by \citet{hegade2023digitized} has also been published, claiming that
a digitized-counterdiabatic quantum computing (DCQC) algorithm outperforms QAOA at the task of 
refining the Babai approximations to the CVP. \citet{hegade2023digitized} present results showing that the DCQC method
retrieves the lowest energy state of the corresponding Hamiltonian, with greater probability than a corresponding QAOA
method. Their report is much shorter than that of \citet{yan2022factoring}, and only compares the two quantum approaches,
without analyzing how this affects the rest of the factoring pipeline. For this reason, our results will be compared with those 
of \citet{yan2022factoring}, which are much more comprehensively explained. Our results will also support the claim that the
improvement claimed by \citet{hegade2023digitized} would not be enough to materially alter our conclusions about whether 
any such variants will enable Schnorr's factoring method to work end-to-end on large numbers.

\section{Factoring Experiments and Results}
\label{sec:experiments}

In this section we discuss the results of various experiments we ran using our own implementation the methods of \citet{schnorr2013factoring} and
\citet{yan2022factoring}. In addition, we developed and implemented alternative lattice-based factoring heuristics that can be understood as variations of Schnorr's original method, and bring extra insight on these claims and results.

The results are summarized in Tables \ref{tab:schnorr_results_table}, \ref{tab:qaoa_results_table} and \ref{tab:local_search_results_table}. 
Each row is an average from factorizing 10 randomly chosen semiprime numbers 
with the given bit-length. The key conclusion is that, while the quantum optimization (QAOA)
obtains more smooth-relation pairs for each lattice tested than the Schnorr method itself, the simple
classical optimization (Local Search) produces an even greater yield using a much easier and faster alternative; the discrepancy between original method and the QAOA optimization is mostly explained by the fact that the latter tests multiple candidates per lattice, while in the former we only check the candidate produced by Babai's nearest plane approximation. In particular, compare the number of smooth candidates each method needed to test in order to find a factor.

\begin{figure}[ht]
    \centering
    \includegraphics[width=0.65\textwidth]{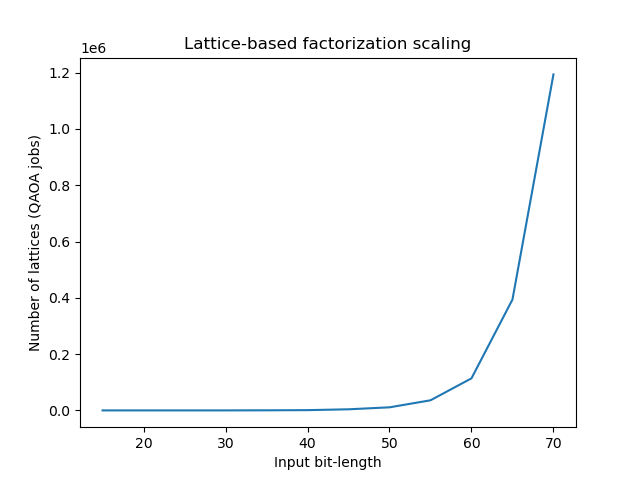}
    \caption{The number of lattices of sublinear dimension needed to factor an $n$-bit integer scales exponentially.}
    \label{fig:lattice_based_fact_scaling}
\end{figure}

It is important to note that the number of lattices required for factorization scales exponentially with the input's bit-length, as the graph in Figure \ref{fig:lattice_based_fact_scaling} suggests: this explains why lattice-based factoring requires exponential time.

The rest of this section analyzes these results and the heuristic alternatives in 
much more detail.

\subsection{Simulation Parameters, Variants, and Detailed Results}

Three configurable hyper-parameters are common to all the heuristics we present below: the lattice dimension $m$ and the so-called ``precision'' parameter $c$ used to define the prime lattice $B_{m, c}$, as in Relation \ref{prime_lattice}; and the length $M$ of the factor basis used to collect $p_M$-smooth candidates. We note while $M = m$ in Schnorr's original proposal, \citet{yan2022factoring} propose using $M > m$ to increase the probability that any given relation pair is smooth. Of course this increases the number of fac-relations that need to be discovered in the collection step, but \citet{yan2022factoring} claim they can reduce the overall computational load with an appropriate choice of $M > m$. However, they provide no guidance on how to select an appropriate $M$ beyond the three concrete examples in Table $(S5)$ that are not accompanied by any justification. 
Hence, the lattice dimension and number of qubits we used do not always match those used by \cite{yan2022factoring} exactly.
Instead, our experiments and results in Table \ref{tab:qaoa_results_table} were configured so that a
repeatable automated process was used for all bitlengths, so that the larger trends are more reliable.

Our implementation makes the following choices by default: given $n$, we set
\begin{equation}
\label{default_params}
    m = \bigg\lceil \frac{3}{2} \frac{\log(n)}{\log(\log(n))} \bigg\rceil,
    \quad
    c = m / 4, 
    \quad \text{and} \quad
    M = m^2.
\end{equation}
Numerical evidence collected from various factoring experiments, much like the ones described below, guided our choice of hyper-parameters, but we make no claims about their optimality. 
Notice our default lattice dimension $m$ is sublinear in the bit-length of $n$, as proposed by \citet{yan2022factoring}.

As an extra check, we ran a 48-bit factorization attempt using a lattice dimension of 10, to follow as closely as possible
the method of \citet{yan2022factoring} for this bit-length. It required more than 95000 lattices to be searched, 
each one of which would be a QAOA optimization job. \citet{yan2022factoring} gloss over the scale of this problem,
saying just ``The calculations of other sr-pairs are similar and will be obtained by numerical method.''
This difference is crucial: it is not factoring a 48-bit number on a quantum computer as claimed, but instead, 
they performed a tiny part of a massively parallel process on a quantum computer.

%\todo[inline]{Run a single 10-qubit QAOA job on Aria-1 to produce plots like those in Figure 4 of \citet{yan2022factoring}. (This will allow us to compare IonQ to IBM %hardware somewhat apples-to-apples. I say somewhat because we may not be able to conduct the same exact experiment with all the same hyperparameters.)}

%%%
\subsection{Heuristic Comparisons and the Availability of Smooth Relations}

A key challenge in predicting the behavior of these methods at scale is in understanding the relationships
between the distribution of smooth numbers and relations, lattice vectors, and shortest vector lengths.
This section describes heuristic methods and experiments that were developed to shed light on these questions.

\begin{figure}[ht]
    \centering
    \includegraphics[width=\linewidth]{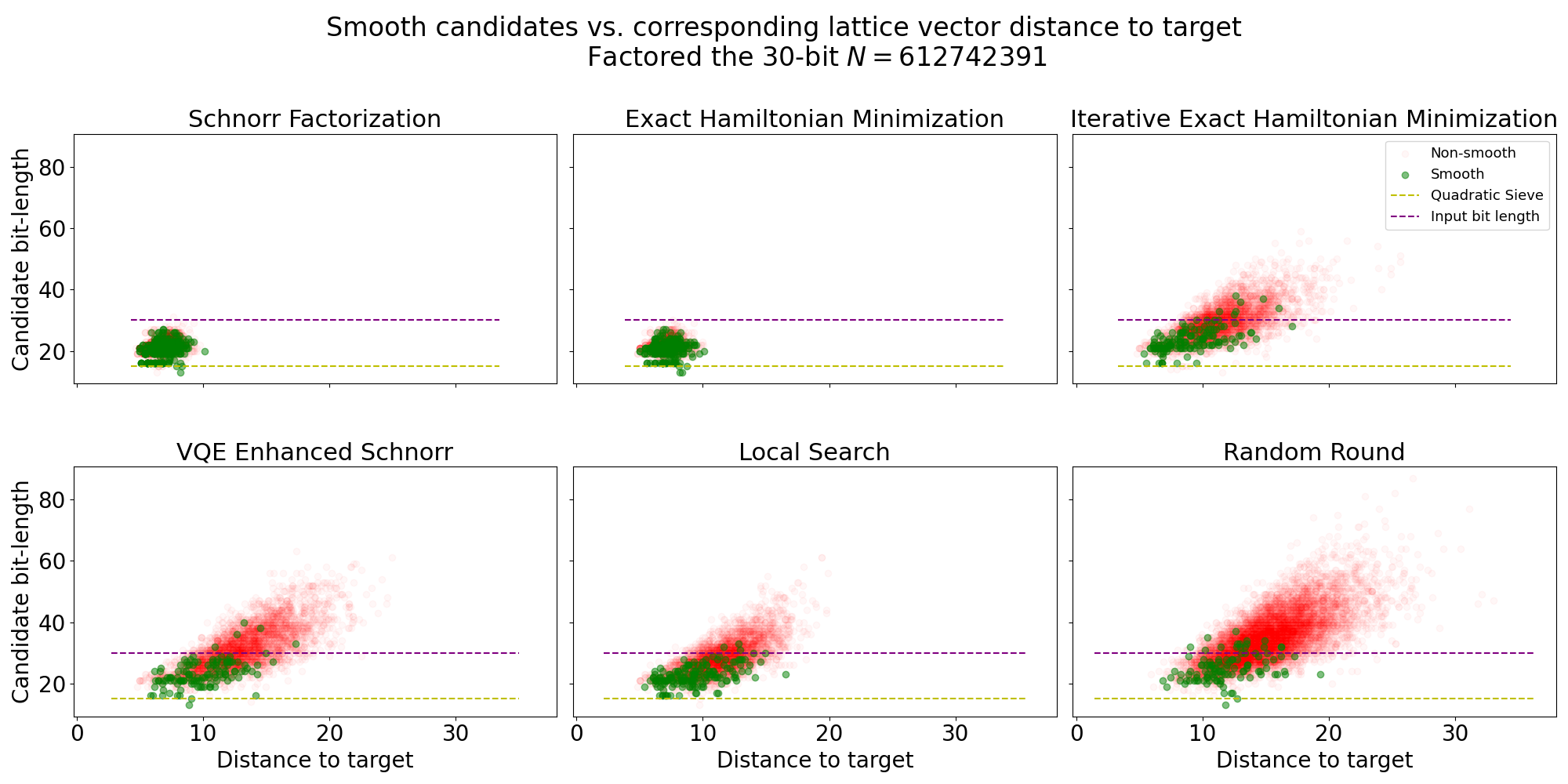}
    \caption{Performance comparison for six different lattice-based heuristics
    for finding smooth relations. The candidate is the number $u-vn$ from which 
    we try to extract a smooth relation. The ideal candidates are smaller (lower down), closer to the target vector (further left). The general trend is clear: candidates with shorter bit-length are more likely to be smooth, as predicted by the Dickman-de Brujin function, and these tend to correspond to better CVP approximations.
    }
    \label{comparison plot}
\end{figure}

\begin{table}
\small
\caption{Schnorr's method: Performance statistics for various bit-lengths}
\label{tab:schnorr_results_table}
\begin{tabular}{rrrrrrrr}
\toprule
\makecell{Input bit \\ length} &  \makecell{Lattice \\ dimension} &  \makecell{Lattices \\ tested} &  
\makecell{Candidates \\ extracted} &
\makecell{Total \\ SR pairs} &  \makecell{Unique \\ SR pairs} &  
\makecell{Unique SR \\ per lattice \%} &  Time (s) \\
\midrule
               15 &                  7 &           490.0 &                490.0 &          352.33 &            37.67 &          8.26 &      0.70 \\
               20 &                  8 &           890.0 &                890.0 &          413.80 &            50.10 &          7.29 &      0.71 \\
               25 &                  9 &          1290.7 &               1290.7 &          335.50 &            75.40 &          6.35 &      0.72 \\
               30 &                 11 &          3046.5 &               3046.5 &          458.20 &            93.40 &          3.27 &      1.33 \\
               35 &                 12 &         11711.6 &              11711.6 &          618.22 &           113.22 &          1.07 &      4.03 \\
               40 &                 13 &         35192.5 &              35192.5 &         1251.80 &           141.90 &          0.43 &     10.36 \\
               45 &                 14 &        220661.0 &             220661.0 &         3090.80 &           175.10 &          0.08 &     73.01 \\
               50 &                 15 &        770965.9 &             770965.9 &         4461.50 &           205.70 &          0.03 &    324.75 \\
               55 &                 16 &       4631264.2 &            4631264.2 &        14339.80 &           240.20 &          0.01 &   2526.92 \\
               60 &                 17 &      15281405.7 &           15281405.7 &        14365.40 &           276.30 &          0.00 &  12528.89 \\
\bottomrule
\end{tabular}

\medskip
\caption{Yan et al. QAOA method: Performance statistics for various bit-lengths}
\label{tab:qaoa_results_table}\centering
\begin{tabular}{rrrrrrrr}
\toprule
\makecell{Input bit \\ length} &  \makecell{Lattice \\ dimension} &  \makecell{Lattices \\ tested} & \makecell{Candidates \\ extracted} &
\makecell{Total \\ SR pairs} &  \makecell{Unique \\ SR pairs} &  \makecell{Unique SR \\ per lattice \%} &  Time (s) \\
\midrule
               15 &                  7 &            91.1 &               1287.2 &          345.6 &           190.1 &        208.66 &      4.66 \\
               20 &                  8 &            98.9 &               1486.8 &          190.0 &           139.9 &        141.48 &      4.89 \\
               25 &                  9 &           179.9 &               2796.8 &          117.0 &           106.7 &         60.63 &     11.23 \\
               30 &                 11 &           469.9 &               7358.8 &          122.7 &           112.5 &         24.09 &     30.83 \\
               35 &                 12 &          1809.6 &              28494.7 &          142.7 &           130.3 &          7.26 &    133.85 \\
%               40 &                 13 &            6700 &                 &           &            &           &    2400 \\
               40 &                 13 &            6200 & 98119.4 & 179.3 & 163.0 & 2.64 & 409.73 \\
%               45 &                 14 &          23900 &               &           &            &           &    4800 \\
               45 &                 14 &           25000 & 397176.9 & 215.8 & 186.5 & 0.753 & 2311.47 \\
% Explicit Yan et al 48 repro, different lattice dimension choices.
% 48 & 10 & 95826 & 1498535 & 1633 & 158 & 0.165 & 9068.97 \\
%               50 &                 15 &          82500  &               &           &            &           &    15305 \\
               50 &                 15 &          88340 & 1406730       & 253.8 & 215.5 & 0.245 & 9495.31 \\
\bottomrule
\end{tabular}

\medskip

\caption{Local Search Method: Performance statistics for various bit-lengths}
\label{tab:local_search_results_table}
\begin{tabular}{rrrrrrrr}
\toprule
\makecell{Input bit \\ length} &  \makecell{Lattice \\ dimension} &  \makecell{Lattices \\ tested} & \makecell{Candidates \\ extracted} &
\makecell{Total \\ SR pairs} &  \makecell{Unique \\ SR pairs} &  \makecell{Unique SR \\ per lattice \%} &  Time (s) \\
\midrule
               15 &                  7 &             92.3 &                1476.8 &           507.2 &            194.5 &        211.13 &      0.73 \\
               20 &                  8 &             97.7 &                1563.2 &           286.7 &            152.2 &        155.85 &      0.66 \\
               25 &                  9 &             99.9 &                1598.4 &           143.1 &            105.0 &        105.10 &      0.63 \\
               30 &                 11 &            220.0 &                3520.0 &           168.4 &            118.6 &         54.97 &      0.95 \\
               35 &                 12 &            570.0 &                9120.0 &           180.4 &            130.0 &         23.18 &      2.08 \\
               40 &                 13 &           1999.9 &               31998.4 &           297.7 &            176.4 &          8.82 &      4.45 \\
               45 &                 14 &           5800 &               92800 &           346.4 &            189.0 &          3.31 &     11.76 \\
               50 &                 15 &          17500 &              280000 &           450.6 &            216.5 &          1.24 &     37.52 \\
               55 &                 16 &          63200 &             1011200 &           615.4 &            244.4 &          0.39 &    176.43 \\
               60 &                 17 &         184500 &             2952000 &           958.7 &            277.3 &          0.15 &    602.54 \\
\bottomrule
\end{tabular}

\end{table}

Figure \ref{comparison plot} compares the performance of six different heuristics used to factor the $30$-bit integer $n = 612742391$ by approximating solutions to the CVP on  lattices of dimension $11$ and using $11^2 = 121$ primes in the factor basis (which thus goes up to $p_{121}=661$). Each plot illustrates the candidates $S = u - vn$ tested for smoothness: there is a green dot for each candidate that turned out to be $p_{121}$-smooth, and a red one for each candidate that was not. The dashed purple line indicates the bit-length of the input $n$, while the dashed yellow line indicates the average bit-length of the candidates for smoothness tested by the Quadratic Sieve.

Notice that in each case the green dots tend to cluster around the lower-left region of the plot. Thus the general trend is clear: candidates with shorter bit-length are more likely to be smooth, as predicted by the Dickman-de Brujin function, and these tend to correspond to better CVP approximations. 

In other words, broadly speaking, lattice vectors that are closer to the target tend to produce candidates that are more likely to be smooth. 
%The plot in Figure \ref{fig:lattice_based_fact_scaling} substantiates this interpretation \todo{Wrong plot! Should we %include the smoothness probability vs distance to target plot too?}. 
In fact Schnorr's claims are based on this tenuous relationship and, as mentioned above, this relationship lays the foundation for \citet{yan2022factoring}'s promise to enhance Schnorr's methods by refining the CVP approximations using quantum computations.

However, the plots in Figure \ref{comparison plot} indicate that the situation is much more complicated. To elucidate this nuance, we must describe each heuristic in some detail. 

We begin with the top row in Figure \ref{comparison plot}. The plot on the left illustrates the smoothness candidates obtained using Schnorr's method together with Babai's nearest plane approximation for the CVP, as described in Section \ref{sec:schnorrgate}. In particular, this heuristic extracts a single smooth candidate from each lattice: first we use Babai's nearest plane algorithm to find a lattice vector close to the target and then we test $S = u - vn$, with $(u, v)$ as defined by Relation \ref{smooth_pair_defn}, for $p_M$-smoothness. (The experiments used the `extended factor basis' for relation search, where an $m$-dimensional lattice
is searched, and relations that are smooth for some $p_M$ where $M > m$ are accepted: this is a departure from strictly 
reimplementing Schnorr's method, but enables more direct comparison with the behavior of the other methods.)

% \todo[inline]{The experiments used the extended factor basis for relations collection... Should we keep as is or re-run without extending the factor basis?}

The plot in the top middle illustrates the candidates obtained by refining Babai's CVP approximation using \textit{exact} Hamiltonian energy minimization. Concretely: first we use Babai's nearest plane algorithm to find a lattice vector close to the target, as above; next we set up the Ising Hamiltonian defined by Relation \ref{babai_hamiltonian}; and then for \textit{each} eigenstate with lower energy than the classical approximation state $\ket{0}$ we test the candidate $S = u - vn$ for $p_M$-smoothness. Again, we take $(u, v)$ as defined by Relation \ref{smooth_pair_defn} but in this case we use the lattice coordinates corresponding to the refined CVP approximation as explained in Relation \ref{lattice_coords}. In essence, this heuristic follows \citet{yan2022factoring}'s proposal except that it obtains the eigenstates exactly using linear algebraic techniques instead of QAOA; in addition, it extracts a smooth candidate from each eigenstate with lower energy than $\ket{0}$, instead of testing lattice coordinates corresponding to the most likely minimum-energy eigenstate candidates as determined by the QAOA. It is important to note that the size of the matrix describing the problem Hamiltonian increases exponentially, so this heuristic may only be used for ``small'' integers: powerful modern laptops can handle inputs with up to $\sim 75$ bits.

The plot on the top right in Figure \ref{comparison plot} illustrates the results for the ``hill-climbing'' heuristic detailed in Algorithm \ref{hill_climb_alg}, with a small modification: if $b^{\mathrm{curr}}$ equals $b^{\mathrm{prev}}$ at the end of the first iteration, we generate a new coding vector by independently and uniformly drawing from $\{0, 1\}$ and try again. Essentially, this heuristic iterates the exact Hamiltonian energy minimization routine described in the last paragraph: at each step, it updates the CVP approximation resulting from the minimum-energy eigenstate. Thus, assuming the hypothesis in \citet{yan2022factoring}, we expect the hill-climbing heuristic to outperform all others because it uses the best CVP refinement out of all the heuristics we propose. 
%This heuristic updates the coding vector according to the coefficients of the minimum-energy eigenvector, in order to allow for changing the search direction in each dimension corresponding to a trivial coefficient. \todo{need to explain the random shaking of the coding vector here...}

\begin{algorithm}[t]
\caption{Hill-climbing refinement heuristic}
\label{hill_climb_alg}
\begin{algorithmic}
\Require prime lattice $B$, LLL-reduction $B^*$, Babai CVP approximation $b^{\mathrm{op}}$, coding vector $\kappa$
\State $b^{\mathrm{prev}} \gets 0, \quad b^{\mathrm{curr}} \gets b^{\mathrm{op}}$
\While{$b^{\mathrm{curr}} \neq b^{\mathrm{prev}}$}
    \State $b^{\mathrm{prev}} \gets b^{\mathrm{curr}}$
    \State Construct Hamiltonian $H$ as in Relation \ref{babai_hamiltonian}
    \State Compute minimum-energy eigenstate $\ket{\psi^*}$
    \State Compute $b^{\ket{\psi^*}}$ as in Relation \ref{cvp_refinement} 
    \State $b^{\mathrm{curr}} \gets b^{\ket{\psi^*}}$
    \For{$j = 1 \ldots m$}
        \If{$\psi_j^* = 1$}
            \State $\kappa_j \gets -\kappa_j$
        \EndIf
    \EndFor
\EndWhile
\State Compute lattice coordinates $e \gets B \backslash b^{\mathrm{curr}}$
\State \Return $S = u - vn$ with $(u, v)$ as defined by Relation \ref{smooth_pair_defn}
\end{algorithmic}
\end{algorithm}

Now we move on to the bottom row of plots. The plot on the bottom left describes the candidates obtained using the proposal described in \citet{yan2022factoring}, with one modification: we use multi-angle QAOA in place of the usual QAOA. We found that assigning an independent parameter to each rotation gate in the variational ansatz allows the quantum optimization routine to obtain eigenstates with lower energy than those obtained with straight-up QAOA. 
This plot displays results computed using IonQ's Aria-1 (noisy) simulator. To produce this plot, we sampled each optimal state distribution obtained using the multi-angle QAOA subroutine 1000 times and we tested up to 16 smooth candidates for each lattice; in this case, each candidate corresponds to one of the eigenstates assigned the highest likelihood by our multi-angle QAOA subroutine via Relation \ref{cvp_refinement}.

The plot in the bottom middle illustrates the results of our \textit{Local Search} heuristic, detailed in Algorithm \ref{local_search}. Essentially, given a search parameter $k$, this heuristic extracts a candidate from each of the $2^k$ possible roundings of the Babai coefficients $c_{1}, \ldots, c_k$, as in Relation \ref{babai_update}, which correspond to the $k$ smallest-norm columns of the LLL-reduced lattice. The plot displays results obtained using $k = 4$ (testing $16$ candidates per lattice) and shows that this heuristic indeed outperforms any claimed quantum advantage as it obtains the factorization of $n$ upon testing less smooth candidates than any other algorithm. Note that the Local Search heuristic does \textit{not} involve any quantum computations.

\begin{algorithm}[t]
\caption{Local search heuristic}
\label{local_search}
\begin{algorithmic}
\Require prime lattice $B$, LLL-reduction $B^*$, Babai CVP approximation $b^{\mathrm{op}}$, search parameter $k$
\State $\texttt{candidates} \gets \mathrm{zeros}(2^k, 1)$
\For{$i = 1 \ldots 2^k$}
    \State Use the binary representation of $i$ to write $\ket{i}$
    \State Construct $b^{\ket{i}}$ as in Relation \ref{cvp_refinement}
    \State Compute lattice coordinates $e \gets B \backslash b^{\mathrm{op}}$
    \State Compute $S = u - vn$ with $(u, v)$ as defined by Relation \ref{smooth_pair_defn}
    \State $\texttt{candidates}[i] \gets S$
\EndFor
\State \Return \texttt{candidates}
\end{algorithmic}
\end{algorithm}

The plot on the right displays the results of our \textit{Random Round} heuristic, which extracted 16 candidates from each lattice, each corresponding to a random rounding of the $m$ Babai coefficients $c_1, \ldots, c_m$, as in Relation \ref{babai_update}, via Relation \ref{cvp_refinement}. We include this plot mainly to demonstrate that the Local Search heuristic chooses roundings intentionally, and these indeed outperform a random selection. 

With a description of the six different factoring heuristics in mind, we are ready for a more nuanced analysis. Though in general it seems that better CVP approximations are more likely to yield smooth relations, it turns out that producing higher quality CVP solutions does not necessarily lead to faster factoring, largely 
because we tend to encounter the same fac-relation over and over: notice that both the Schnorr and the exact Hamiltonian minimization heuristic see the same fac-relation more than three times on average; notice that the same ratio is much closer to 1 when using any of the heuristics in the bottom row. This observation undermines \citet{yan2022factoring}'s argument: while we may obtain better CVP approximations in some cases using quantum computations, these do not yield enough unique fac-relations sooner. Thus we observe no quantum advantage when following the method proposed in \citet{yan2022factoring}, even when we improve it by replacing QAOA with its multi-angle variant.

\begin{figure}[!t]
    \centering
    \includegraphics[width=4in]{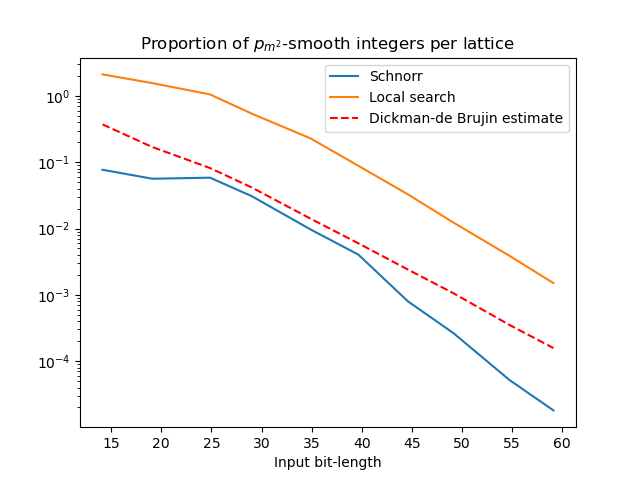}
    \caption{The proportion of fac-relations extracted per lattice tested decreases exponentially when the lattice dimension $m = \lceil \frac{3}{2} \frac{\log(n)}{\log(\log(n))} \rceil$ is sublinear in the bit-length of $n$, as proposed in \citet{yan2022factoring}.}
    \label{fig:smooth_ints_prop}
\end{figure}

A further key point to consider is that \citet{yan2022factoring} vastly underestimate the number of qubits needed to reliably unique collect fac-relations. The number of qubits in \citet{yan2022factoring}'s proposal is equal to the dimension of the lattice used. The plot in Figure \ref{fig:smooth_ints_prop} shows that when the lattice dimension $m = \lceil \frac{3}{2} \frac{\log(n)}{\log(\log(n))} \rceil$ is sublinear in the bit-length of $n$, as proposed in \citet{yan2022factoring}, the proportion of fac-relations extracted per lattice  decreases exponentially. This means that factoring with sublinear resources still takes exponential time, because we need to test exponentially many lattices in order to collect enough fac-relations.

\subsection{Quantum Processor Optimization for QAOA Jobs}

In this section we discuss the results of the QAOA experiments we conducted, in particular
following the 3-qubit case of \citet{yan2022factoring}.
In this case we seek to factor $n = 1961$ using lattices of dimension $m = 3$. According to Table $(S5)$ in \citet{yan2022factoring}, the factor basis has $15$ primes, so we extract smooth relation candidates using lattices of dimension $m = 3$ and collect all pairs that are $p_{15}$-, or $47$-smooth. As in \citet{yan2022factoring}, we fix $c = 1.5$. In addition, we consider the prime lattice 
\begin{equation*}
    B_{3, \, 1.5} = 
\left[\begin{array}{rrr}
1 & 0 & 0 \\
0 & 1 & 0 \\
0 & 0 & 2 \\
22 & 35 & 51
\end{array}\right]
\end{equation*}
and the target vector
\begin{equation*}
    t = \left[0,\,0,\,0,\,240\right],
\end{equation*}
as in Relation $(S37)$ of \citet{yan2022factoring}. Applying the LLL-reduction algorithm, with parameter $\delta = 0.99$, to the prime lattice $B_{3, \, 1.5}$  yields the reduced matrix
\begin{equation*}
    B^* = 
\left[\begin{array}{rrr}
1 & -4 & -3 \\
-2 & 1 & 2 \\
2 & 2 & 0 \\
3 & -2 & 4
\end{array}\right],
\end{equation*}
as in Relation $(S40)$ of \citet{yan2022factoring}.\footnote{Here we take a moment to note a discrepancy in \citet{yan2022factoring}: while Algorithm $2$ claims they compute LLL reductions with parameter $\delta = 0.75$, the {\sc SageMath} implementation of the LLL-reduction algorithm with this parameter yields
\begin{equation*}
\left[\begin{array}{rrr}
1 & -3 & -4 \\
-2 & 2 & 1 \\
2 & 0 & 2 \\
3 & 4 & -2
\end{array}\right]
\end{equation*}
instead.}

When we apply Babai's method to approximate the closest lattice vector to $t$, using the reduced matrix $B^*$, we obtain
\begin{equation*}
    b^{\mathrm{op}} = 
    \left[0,\,4,\,4,\,242\right]^T
    \quad\text{and}\quad
    \kappa = 
    \left[-1,\,-1,\,-1\right]
\end{equation*}
as in Relation $(S44)$ and Table $(S3)$ in \citet{yan2022factoring}. 

Using Relation \ref{babai_hamiltonian}, we obtain the Hamiltonian
\[H = -4 Z_{0} Z_{1} + \frac{5}{2} Z_{0} Z_{2} + 3 Z_{1} Z_{2} - \frac{3}{2} Z_{0} - \frac{7}{2} Z_{1} - 4 Z_{2} + \frac{87}{2},
\]
as in Relation $(S52)$ in \citet{yan2022factoring}. Table \ref{tab:3_qubit_fac_rels} describes the eigen-pairs and corresponding relation pairs associated to the lowest $4$ energies of $H$.

\begin{table}[ht]
    \centering
    \begin{tabular}{c|c|c|c|c|c}
        Energy Level &  Eigenvalue & Eigenstate & $(u, v)$ & $S = u - vN$ & Is $p_{15}$-smooth? \\
        \hline
        \hline
        0 & 33 & 100 & (1800, 1) & $-7 \times 23$ & Yes \\
        1 & 35 & 011 & (1944, 1) & $-17$ & Yes \\
        2 & 36 & 000 & (2025, 1) & $2^6$ & Yes \\
        3 & 42 & 001 & (3645, 2) & 277 & No \\
    \end{tabular}
    \caption{The eigen-pairs and corresponding relation pairs associated to the lowest $4$ energies of $H$.}
    \label{tab:3_qubit_fac_rels}
\end{table}

\begin{figure}[ht]
    \centering
    \includegraphics[width=\textwidth]{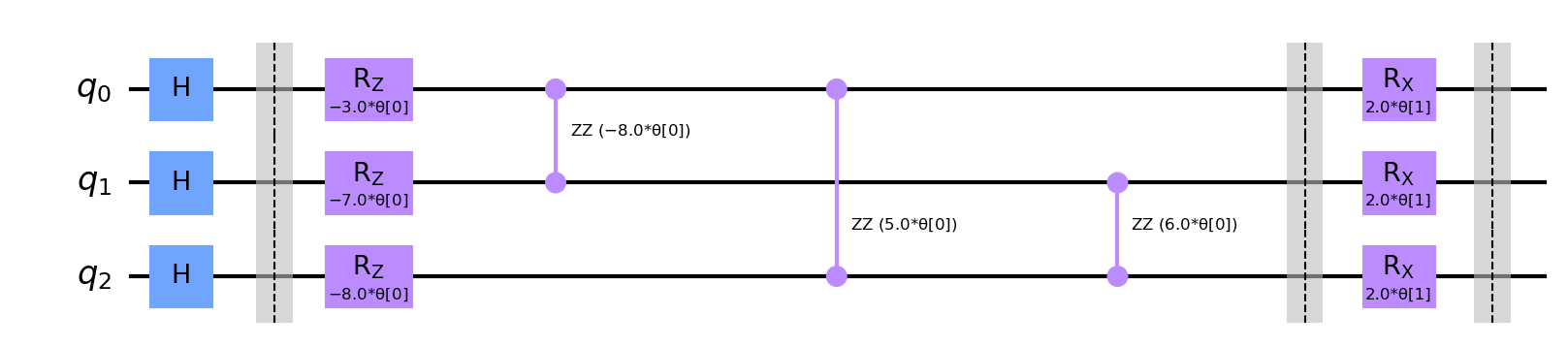}
    \caption{QAOA circuit for 3-qubit Hamiltonian, single layer (depth $p=1$)}
    \label{fig:qaoa_circuit}
\end{figure}

Then Figure \ref{fig:qaoa_circuit} illustrates a single layer of the quantum circuit used in the associated QAOA computation.

\begin{figure}[ht]
    \centering
    % Note the use of trim=2cm 0 2cm 0,clip to reduce image margins, thus enlarging the useful space in the image. 
    \includegraphics[trim=2cm 0 2cm 0,clip,width=0.9\textwidth]{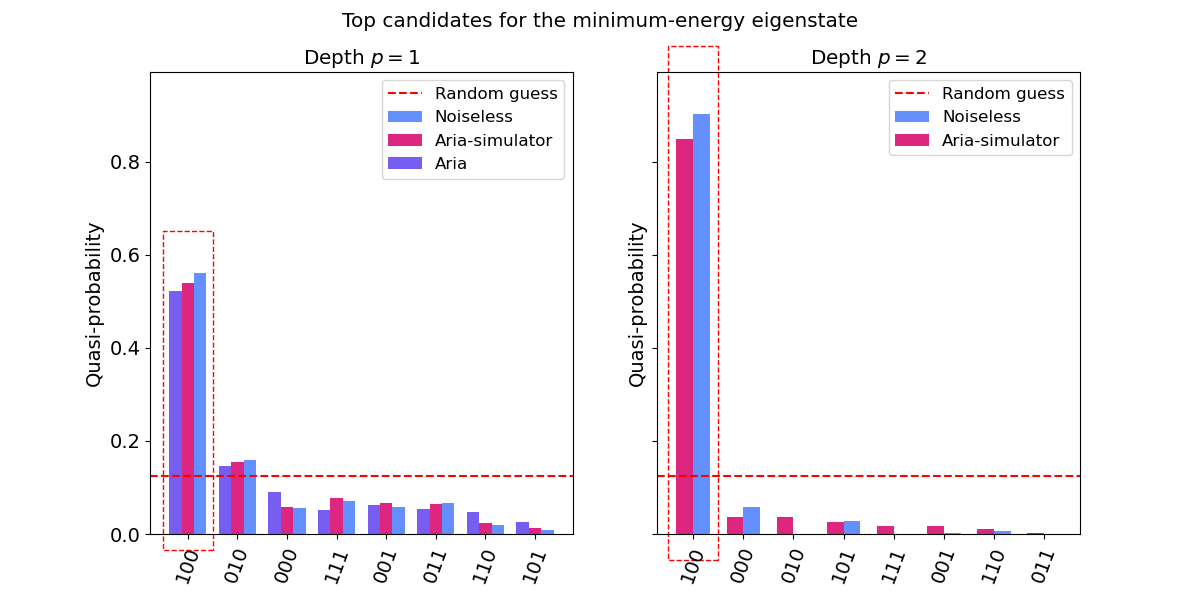}
    % \caption{Results of 3-qubit QAOA job with depth $p=1$}
    % \label{fig:3_qubit_depth_1_results}

    \centering
    \includegraphics[trim=2cm 0 2cm 0,clip,width=0.9\textwidth]{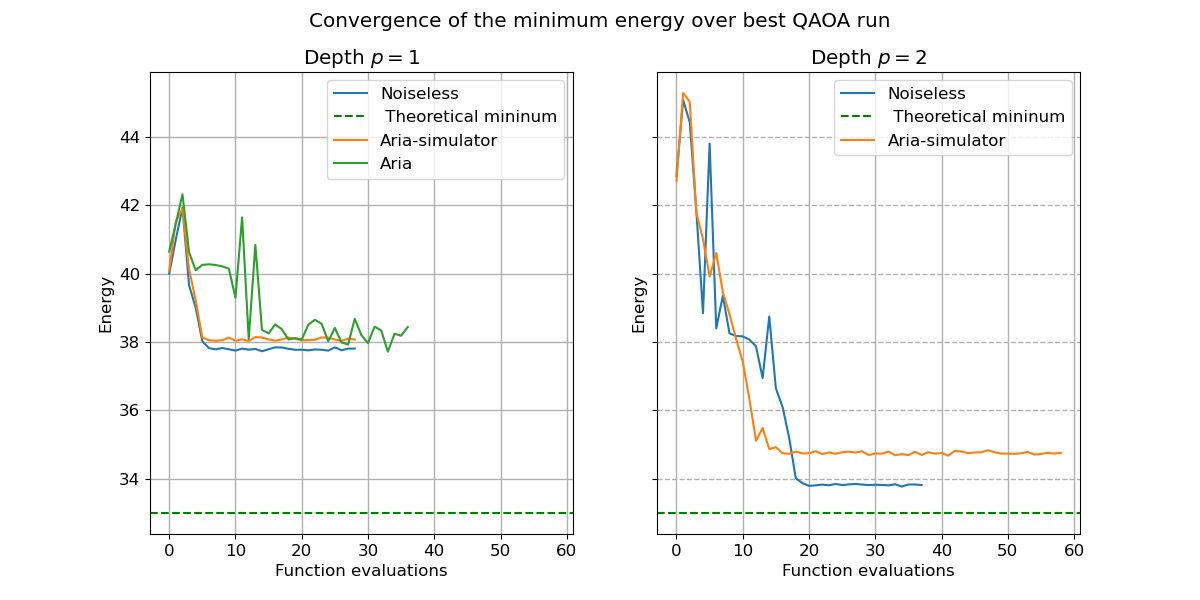}
    \caption{Results of 3-qubit QAOA job with depths $p=1, 2$. In the depth $p=1$ case, we ran a full QAOA calculation using the classical BFGS optimization routine starting at the same initial value on three different backends: the noiseless Qiskit Aer simulator, the IonQ Aria simulator, and finally on the IonQ Aria QPU. The results show good agreement between the simulators and the quantum hardware in this toy example.}
    \label{fig:3_qubits_qaoa_results}

    % \centering
    % \includegraphics[trim=2cm 0 2cm 0,clip,width=\textwidth]{images/qaoa_results_3_3_.png}
    % \caption{Results of 3-qubit QAOA job with depth $p=3$}
    % \label{fig:3_qubit_depth_3_results}
\end{figure}

Figure \ref{fig:3_qubits_qaoa_results} plots the results of QAOA experiments conducted using quantum circuits with $p = 1$ and $p = 2$ layers, contrasting
noiseless and noisy simulation. In addition, in the depth $p=1$ case we executed the QAOA run on IonQ's Aria QPU. Our implementation features expected energy calculations that are differentiable with respect to the lattice parameters, so we use the quasi-Newton BFGS optimization routine.
We rely on \texttt{SciPy}'s implementation and leverage analytical gradient evaluations using our implementation of a parameter-shift rule. For each depth $p$, every execution uses the same initial parameter values. These parameter values were chosen amongst $50$ sets of initial values used in ``dry runs'' of the experiment, where we relied on Qiskit's noiseless Aer simulator. Each set of initial values in the ``dry runs'' was generated uniformly at random from $[-\pi, \pi]$.

The results show good agreement between the simulators and the hardware in this simple example. We note that we do see improved convergence in the depth $p=2$ case as we increase the number of circuit layers, but the noisy and noiseless simulations diverge further. Though indeed in theory ``the quality of the optimization improves as $p$ is increased'', as stated in \citet{farhi2014quantum}, in practice we observe that adding too many more layers can reduce the quality of the solution, because the additional noise incurred by running a deeper circuit eventually outweighs the gains.

% All but the noiseless $p=3$ results put the states $\ket{100}$ and $\ket{011}$ correctly in the first two positions.
% Only the noisy $p=3$ result correctly put $\ket{000}$ in third place, which is the other eigenstate that leads to a smooth factoring relation.

% if we have something interesting to say we'll include these results.
% \subsection{5-qubit case}

% \begin{figure}[ht]
%     \centering
%     \includegraphics[width=\textwidth]{images/qaoa_results_5_2_.png}
%     \caption{Results of 5-qubit QAOA job with depth $p=2$}
% \end{figure}

% \subsection{10-qubit case}

% %%%%
% \begin{figure}[ht]
%     \centering
%     \includegraphics[width=\textwidth]{images/qaoa_results_10_1_.png}
%     \caption{Results of 10-qubit QAOA job with depth $p=1$}
% \end{figure}

% {\it Run jobs that reproduce \citet[\S3]{yan2022factoring}, while commenting on how these aren't really useful results. All we want
% to show is that our diff between theory and experiment is smaller.}

%%
\section{Further Lattice Variations Attempted}

As well as the different heuristic techniques whose results are documented above, other attempts to improve factorization 
performance included:

\begin{itemize}
    \item Varying the strategy for choosing lattices to search. \cite{schnorr2021fast} and \cite{yan2022factoring} use slightly
    different lattice basis vectors, both derived from random permutations. We explored these options, and sampling from other
    distributions. This did not significantly improve results.
    \item Redundancy (finding the same smooth relation from many lattices) was a key problem throughout. We tried to 
    introduce deliberate variation by choosing lattices from permutations with large $\ell_1$-norm distances from permutations already used:
    there was no improvement in the rate of finding new smooth relations (and a large computational cost). 
\end{itemize}

While it was infeasible to try every combination of options, these experiments further demonstrated that there is
no clear path to factoring large numbers using this family of methods.

%%%
\section{Possible Alternatives for Quantum Advantage}

The closest vector problem is related to the knapsack problem \citep{salkin1975knapsack}. 
In the knapsack problem, the challenge
is to find a combination of objects of different sizes or weights that can fit into a knapsack with a given 
carrying capacity: so if we  
add the restriction that all the coefficients
$e_j$ in Equation \ref{log_approx} must be positive, and the sum must not be greater 
than $\ln n$, this effectively transforms the CVP into a knapsack problem.
We were able to use similar optimization methods to obtain good results for knapsack problems and related variants, using variational quantum
circuits. These results will be described in further work.

We also considered the use of quantum computing for the data processing part of the factoring pipeline. 
The data processing part consists of solving a linear system of equations modulo 2. As well as factoring,
modulo-2 linear systems have applications in error correction and cryptography. 
This potential application of quantum computing to a different part of the factoring problem is documented
by \citet{aboumrad2023modulo2}. 

%%%
% \input{implementation}

%%%
\section{Conclusion}

We have carefully reviewed the methods and claims of \citet{schnorr2021fast} and \citet{yan2022factoring}, which
led to speculation that lattice-based factorization of large numbers could be tractably achieved using classical 
or near-term quantum computers. We also implemented these methods in a complete factorization pipeline, which supports
a much more systematic analysis of the computational claims in practice.

In spite of its many interesting mathematical properties, Schnorr's method does not lead to a faster factoring algorithm
than the sieving methods already available in standard libraries. Optimizations of Schnorr's method can reduce the number of
lattices we need to test, but do not alter the fundamental problem, which is that smooth relations are rare and hard to find.
The chief advantage of QAOA over Schnorr's method is because QAOA tests many smooth relation candidates per lattice: 
but this multiplicity can easily be tried with other methods, with greater improvements, as shown with the Random Round and Local Search
heuristics introduced in this paper.

Though \citet{yan2022factoring} may be correct in their assertion that parts of a factorization calculation 
of a 2048-bit integer may `fit' in a
QPU with 372 qubits (in the sense that one could run the QAOA job to enhance the Babai CVP approximation), it would still take 
exponential time to complete the factorization because the probability of actually finding a fac-relation is exponentially small. In other 
words, while it would be possible to run the appropriate QAOA jobs, we would need to run exponentially many of them in order to 
factorize the input. This means that RSA is still safe from this kind of attack.

When analyzing claims of quantum advantage, it is important to consider systems as a whole. In this case, the maxim that 
``a chain is as strong as its weakest link'' suggests an analysis which revealed clear flaws in Schnorr's method. Instead, 
starting from the principle that ``a chain is as interesting and its most interesting link'' has led to considerable confusion
in the community, because a small quantum optimization in one part of a process can be used to claim an overall 
quantum advantage, for which there is no end-to-end evidence.

Nonetheless, there are potentially useful directions for quantum computing in these areas. Of these, we believe the use of 
quantum computers to solve modulo 2 linear systems of equations to be promising.

%%%

\small
\bibliography{ionq}
\bibliographystyle{plainnat}

\end{document}